\documentclass[submission]{eptcs}

\providecommand{\event}{Quantum Physics and Logic 2017}

\usepackage{tikz} 
\usepackage{physics}
\usepackage{amsfonts}
\usepackage{amsthm}
\usepackage{amsmath}
\usepackage{algorithm}
\usepackage{algorithmic}
\usepackage{enumerate}
\usepackage{enumitem}
\usepackage{amssymb}
\usepackage{underscore}
\usepackage{cite}

\newtheorem{theorem}{Theorem}[section]

\newtheorem{lemma}{Lemma}[section]
\newtheorem{definition}{Definition}[section]

\newcommand{\ANNASCOMMENT}[1]{}

\newcounter{list}
\stepcounter{list}

\title{Information Theoretically Secure Hypothesis Test for Temporally Unstructured Quantum Computation \\ \large (Extended Abstract)}

\author{
Daniel Mills
\institute{School of Informatics, University of Edinburgh, UK}
\email{daniel.mills@ed.ac.uk}
\and
Anna Pappa
\institute{School of Informatics, University of Edinburgh,UK}
\institute{Department of Physics, University College London, UK}
\email{annapappa@gmail.com}
\and
Theodoros Kapourniotis
\institute{School of Informatics, University of Edinburgh, UK}
\institute{Department of Physics, University of Warwick, UK}
\email{T.Kapourniotis@warwick.ac.uk}
\and
Elham Kashefi
\institute{School of Informatics, University of Edinburgh, UK}
\institute{LIP6, CNRS, Pierre et Marie Curie University, France}
\email{ekashefi@inf.ed.ac.uk}
}

\def\titlerunning{Information Theoretically Secure Hypothesis Test for the IQP Machine}
\def\authorrunning{D.Mills, A.Pappa, T.Kapourniotis \& E.Kashefi}

\begin{document}

	\maketitle
	
	\begin{abstract}			
        The efficient certification of classically intractable quantum devices has been a central research question for some time. However, to observe a ``quantum advantage'', it is believed that one does not need to build a large scale universal quantum computer, a task which has proven extremely challenging. Intermediate quantum models that are easier to implement, but which also exhibit this quantum advantage over classical computers, have been proposed. In this work, we present a certification technique for such a sub-universal quantum server which only performs commuting gates and requires very limited quantum memory. By allowing a verifying client to manipulate single qubits, we exploit properties of measurement based blind quantum computing to give them the tools to test the ``quantum superiority'' of the server.
	\end{abstract}
	
\section{Introduction}
\label{sec:introduction}

Quantum computers are believed to be able to efficiently simulate quantum systems \cite{feynman1982simulating,georgescu2014quantum} and outperform classical systems for specific tasks. Examples include Shor's algorithm for prime factorisation \cite{shor1999polynomial}, Grover's algorithm for unstructured search \cite{grover1996fast}, and the BB84 \cite{bennett1984quantum} and Ekert91 \cite{ekert1991quantum} protocols for public key exchange. That said, it may be some time before a large scale universal quantum computer capable of demonstrating the computational power of these protocols is built. In the meantime, several intermediate, non-universal models of quantum computation, which are still believed to not be classically simulatable, may prove easier to implement. Examples include the one clean qubit model \cite{knill1998power,morimae2014hardness}, the boson sampling model \cite{aaronson2011computational,gard2015introduction,neville2017no} and the Ising model \cite{gao2017quantum,van2008completeness}. The \emph{Instantaneous Quantum Poly-time} (IQP) machine \cite{shepherd2009temporally} is another such non-universal model with significant practical advantages \cite{aliferis2009fault, bermejo2017architectures}. IQP uses only commuting gates but is believed to remain hard to classically simulate \cite{bermejo2017architectures,bremner2010classical,bremner2016average} even in a noisy environment \cite{bremner2016achieving, bermejo2017architectures}. Providing evidence that a machine can perform hard IQP computations would be a proof of \emph{quantum superiority} \cite{preskill2012quantum} before a universal quantum computer has been realised experimentally.

In \cite{shepherd2009temporally}, the authors present a \emph{hypothesis test} which can be passed only by devices capable of performing hard IQP computations. In order to accommodate a purely classical client, computational assumptions (conjecturing the hardness of finding hidden sub-matroids) are required in order to prove quantum superiority. In the present work, by endowing the client with the ability to perform simple qubit manipulations similar to those used in Quantum Key Distribution schemes \cite{bennett1984quantum}, we develop an information-theoretically secure hypothesis test for IQP. 

The remainder of the paper proceeds as follows. In Section \ref{sec:Preliminaries}, we formally introduce the IQP machine and provide an implementation in Measurement Based Quantum Computing (MBQC) \cite{raussendorf2001one,raussendorf2003measurement} which is more suitable  for proving security in our framework than previous ones \cite{shepherd2009temporally,hoban2014measurement}. In Section \ref{sec:Blind IQP} we use tools from blind quantum computing \cite{broadbent2009universal,fitzsimons2012unconditionally} to derive a delegated protocol for IQP computations which keeps the details of the computation hidden from the device performing it. We prove information-theoretical security of that scheme in a composable framework. Finally, in Section \ref{sec:hypothesis test} we develop a hypothesis test that a limited quantum client can run to certify the quantum superiority of an untrusted Server. 


\section{Preliminaries}
\label{sec:Preliminaries}
\subsection{X-programs}
\label{subsec:Xprograms}

The IQP machine introduced in \cite{shepherd2009temporally} is defined by its capacity to implement $X$-programs. 
\begin{definition}
	An \emph{$X$-program} consists of a Hamiltonian comprised of a sum of products of $X$ operators on different qubits, and $\theta\in[0,2\pi]$ describing the time for which it is applied.  The $h^{th}$ term of the sum has a corresponding vector $\mathbf{q}_{h}\in \left\{ 0 , 1 \right\} ^{n_{p}}$, called a \emph{program element}, which defines on which of the $n_p$ input qubits, the product of $X$ operators which constitute that term, acts. The vector $\mathbf{q}_{h}$ has 1 in the $j$-th position when $X$ is applied on the $j$-th qubit.
    
	As such, we can describe the $X$-program using $\theta$ and the matrix $\mathbf{Q}=(\mathbf{Q}_{hj})\in\{0,1\}^{n_a\times n_p}$ which has as rows the program elements $\mathbf{q}_{h}$, $h=1,\dots,n_a$.
\end{definition}
\vspace{0.1in}
Applying the $X$-program discussed above to the computational basis state $\ket{0^{n_p}}$ and measuring the result in the computational basis can also be viewed as a quantum circuit with input $\ket{0^{n_p}}$, comprised of gates diagonal in the Pauli-X basis, and classical output. Using the random variable $\mathtt{X}$ to represent the distribution of output samples, the probability distribution of outcomes $\widetilde{x} \in \{0,1\}^{n_{p}}$ is:
\begin{equation}
	\label{equ:IQP probability distribution}
	\mathbb{P} \left( \mathtt{X} = \widetilde{x} \right) = \left| \bra{\widetilde{x}} \exp \left( \sum_{h=1}^{n_a} i \theta \bigotimes_{j: \mathbf{Q}_{hj} = 1} X_{j}\right) \ket{0^{n_{p}}}  \right|^2
\end{equation}

\begin{definition}
	\label{def:IQP oracle}
	Given some X-program, an \emph{IQP machine} is any computational method capable of efficiently returning a sample $\widetilde{x} \in \{0,1\}^{n_{p}}$ from the probability distribution \eqref{equ:IQP probability distribution}.
\end{definition}


\subsection{IQP In MBQC}
\label{subsec:IQP in MBQC}

A common framework for studying quantum computation is the MBQC model \cite{raussendorf2001one, raussendorf2003measurement}, where a quantum operation is expressed by a set of measurement angles on an entangled state described by a graph. This entangled state is built by applying a controlled-$Z$ operation between qubits when there is an edge in the corresponding graph. The probabilistic nature of the measurements of the qubits in this state introduces some randomness which may  be corrected for by adjusting the angle of measurement of subsequent qubits depending on the outcomes of the already performed measurements. The entangling, measuring and correcting operations on a set of qubits are usually referred to as a \emph{measurement pattern} \cite{danos2006determinism,danos2007measurement}.

In this work, we will deal with a specific type of graphs, given in the following definition:

\begin{definition}
\label{def:IQP graph}
We define the \emph{IQP graph} of an $X$-program $(\mathbf{Q},\theta)$, as the graph with biadjacency matrix $\mathbf{Q}=(\mathbf{Q}_{hj})\in\{0,1\}^{n_a\times n_p}$. This means that there is a bipartition of vertices into two sets $P$ and $A$ of cardinality $n_p$ and $n_a$ and that an edge exists in the graph between vertex $a_h$ of set $A$ and vertex $p_j$ of set $P$ when $\mathbf{Q}_{hj}=1$.
\end{definition}

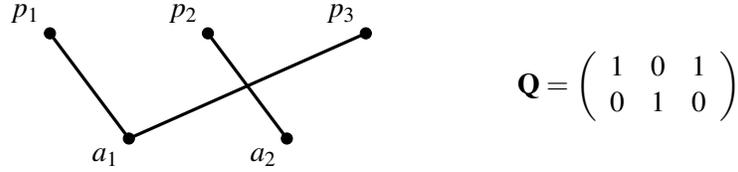
\begin{figure}[!ht]
	\centering
	\begin{tikzpicture}[scale = 0.7]
		\filldraw (1.5,0) circle (3pt) node[anchor = north east] {$a_{1}$};
		\filldraw (4.5,0) circle (3pt) node[anchor = north east] {$a_{2}$};
		
		\filldraw (0,2) circle (3pt) node[anchor = south east] {$p_{1}$};
		\draw[very thick] (0,2) -- (1.5,0);
		
		\filldraw (3,2) circle (3pt) node[anchor = south east] {$p_{2}$};
		\draw[very thick] (3,2) -- (4.5,0);
		
		\filldraw (6,2) circle (3pt) node[anchor = south east] {$p_{3}$};
		\draw[very thick] (6,2) -- (1.5,0);
		
		\node at (11,1) {$\mathbf{Q} =   \left( 
		                                    \begin{array}{ccc}
                                                1 & 0 & 1 \\
                                                0 & 1 & 0
                                            \end{array}
                                        \right)$};
	\end{tikzpicture}
	\caption{An example of an IQP graph described by matrix $\mathbf{Q}$. Here, $n_{p} = 3$ and $n_{a} = 2$ while the partition used is $P = \left[ p_1 , p_2 , p_3 \right]$ and $A = \left[ a_1 , a_2 \right]$.}
	\label{fig:bipartite graph}
\end{figure}

In what follows, we will denote with $\mathbf{Q}$ both the matrix of the $X$-program and its corresponding IQP graph (see Figure \ref{fig:bipartite graph} for an example).  The sets of vertices $A = \left\{ a_{1} , ... , a_{n_{a}} \right\}$ and $P = \left\{ p_{1} , ... , p_{n_{p}} \right\}$ will be called \emph{primary} and \emph{ancillary vertices} respectively. A result which is vital to the remainder of this paper, is the following:

\begin{lemma}
	\label{lem:IQP graph design}
	A measurement pattern can always be designed to simulate an $X$-program efficiently.
\end{lemma}

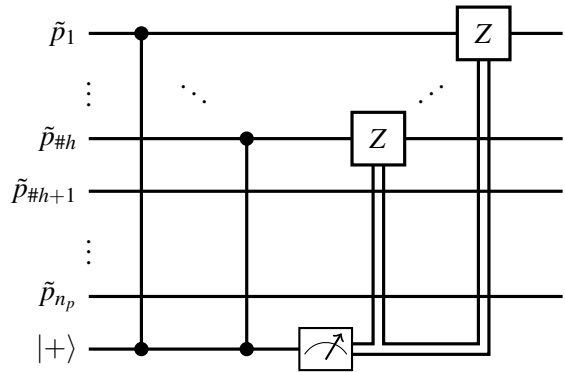
\begin{figure}[!ht]
	\centering
	\begin{tikzpicture}[scale = 0.7]
		\draw[very thick] (0,6) node[anchor=east] {$\tilde{p}_1$} -- (7,6);
		\node at (0,5) {$\vdots$};
		\draw[very thick] (0,4) node[anchor=east] {$\tilde{p}_{\# h}$} -- (5,4);
		\draw[very thick] (0,3) node[anchor=east] {$\tilde{p}_{\# h+1}$} -- (9,3);
		\node at (0,2) {$\vdots$};
		\draw[very thick] (0,1) node[anchor=east] {$\tilde{p}_{n_{p}}$} -- (9,1);
		\draw[very thick] (0,0) node[anchor=east] {$\ket{+}$} -- (4,0);
		
		\filldraw[very thick] (1,0) circle (3pt) -- (1,6) circle (3pt);
		
		\node at (2,5) {$\ddots$};
		
		\filldraw[very thick] (3,0) circle (3pt) -- (3,4) circle (3pt);
		
		\draw[thick] (4,-0.4) rectangle (5,0.4);
		\draw (4.1,-0.1) .. controls (4.3,0.2) and (4.7,0.2) .. (4.9,-0.1);
		\draw[thick, ->] (4.5, -0.2) -- (4.8, 0.3);
		
		\draw[very thick] (5,-0.1) -- (7.6,-0.1) -- (7.6,5.5);
		\draw[very thick] (5,0.1) -- (5.4,0.1) -- (5.4,3.5);
		\draw[very thick] (5.6,3.5) -- (5.6,0.1) -- (7.4,0.1) -- (7.4,5.5);
		
		\draw[very thick] (5,3.5) rectangle (6,4.5) node[pos = 0.5] {$Z$};
		\draw[very thick] (6,4) -- (9,4);
		
		\node at (6.5,5) {\reflectbox{$\ddots$}};
		
		\draw[very thick] (7,5.5) rectangle (8,6.5) node[pos = 0.5] {$Z$};
		\draw[very thick] (8,6) -- (9,6);
		
	\end{tikzpicture}
	\caption{The circuit implementing one term in the sum of equation \eqref{equ:IQP probability distribution}. The input qubits $\{p_j\}_{j=1}^{n_p}$ are rearranged so that if $\# h$ is the Hamming weight of row $h$ of matrix $\mathbf{Q}$, then for $k=1,\dots,\#h$ each $\tilde{p}_k$ corresponds to one $p_j$ such that $\mathbf{Q}_{hj}=1$ and for $k=\#h+1,\dots,n_p$ they correspond to the ones such that $\mathbf{Q}_{hj}=0$.  The ancillary qubit measurement is in the basis $\left\{ \ket{0_{\theta}} , \ket{1_{\theta}} \right\}$ defined in expression \eqref{equ:IQP measuremnt basis}.}
	\label{fig:z-prog circuit}
\end{figure}

One can prove \cite{mills2017information} that the distribution of \eqref{equ:IQP probability distribution} may be achieved by initialising $n_p$ \emph{primary qubits} in the states $\ket{p_{j}} = \ket{+}$, $n_a$ \emph{ancillary qubits} in the states $\ket{a_{h}} = \ket{+}$, applying Controlled-$Z$ operations between qubits when there is an edge in the IQP graph described by the $X$-program matrix $\mathbf{Q}$ and measuring the resulting state. We form this proof by demonstrating that sampling from the distribution in equation \eqref{equ:IQP probability distribution} is equivalent to inputting the primary qubits into a circuit made from composing circuits like the one in Figure \ref{fig:z-prog circuit} (one for each term of the sum in equation \eqref{equ:IQP probability distribution}) and measuring the result in the Hadamard basis. We then argue that all measurements may be delayed to the end of the composed circuit while the ancillary measurement basis is:

\begin{equation}
	\label{equ:IQP measuremnt basis}
	\left\{ \ket{0_{\theta}} , \ket{1_{\theta}} \right\}= \left\{ \frac{1}{\sqrt{2}} \left(  e^{-i\theta}\ket{+} + e^{i\theta}\ket{-} \right) ,\frac{1}{\sqrt{2}} \left(e^{-i\theta} \ket{+} - e^{i\theta}\ket{-} \right)  \right\} 
\end{equation}

\section{Blind Delegated IQP Computation}
\label{sec:Blind IQP}

We now move to build a method for blindly performing an IQP computation in a delegated setting. We consider a client with limited quantum power delegating an IQP computation to a powerful Server. The novel method that we use in this work is to keep the $X$-program secret by concealing the quantum state used. Intuitively, this is done by the client asking the Server to produce a quite general quantum state and then `move' from that one to the one that is required for the computation. If this is done in a blind way then the Server only has knowledge of the general starting state from which any number of other quantum states may have been built. Hence, there are three key problems to be addressed:

\begin{enumerate}
	\item \label{pt:BIQP problem 1} How to transform one state to another.
	\item \label{pt:BIQP problem 2} Which general quantum state to build and transform into the one required for the IQP computation.
	\item \label{pt:BIQP problem 3} How to do this transformation secretly in a delegated setting.
\end{enumerate}


\subsection{Break and Bridge}
\label{sec:Break, Bridge Operators}

The break and bridge operations \cite{fitzsimons2012unconditionally, hein2004multiparty} on a graph $\widetilde{G}=(\widetilde{V},\widetilde{E})$, with vertex set $\widetilde{V}$ and edge set $\widetilde{E}$ describe the operations necessary to solve problem \ref{pt:BIQP problem 1}. 

\begin{definition}
	\label{def:bridge and break}
	The \emph{break} operator acts on a vertex $v \in \widetilde{V}$ of degree 2 in a graph $\widetilde{G}$. It removes $v$ from $\widetilde{V}$ and also removes any edges connected to $v$ from $\widetilde{E}$. 
  
	The \emph{bridge} operator also acts on $v \in \widetilde{V}$ of degree 2 in a graph $\widetilde{G}$. It removes $v$ from $\widetilde{V}$, removes edges connected to $v$ from $\widetilde{E}$ and adds an edge between the neighbours of $v$.
\end{definition}

\begin{figure}[!h]
	\centering
	\begin{tikzpicture}[scale = 0.7]
		\filldraw (0,0) circle (3pt);
		\filldraw (0,1) circle (3pt);
		\filldraw (0,2) circle (3pt);
		\filldraw (1,0) circle (3pt);
		\filldraw (1,1) circle (3pt);
		\filldraw (1,2) circle (3pt);
		\filldraw (2,0) circle (3pt);
		\filldraw (2,1) circle (3pt);
		\filldraw (2,2) circle (3pt);
		
		\draw[very thick] (0,0) -- (0,1);
		\draw[very thick] (0,1) -- (1,0);
		\draw[very thick] (2,0) -- (2,1);
		\draw[very thick] (2,1) -- (2,2);
		\draw[very thick] (0,2) -- (1,2);
		\draw[very thick] (1,0) -- (1,1);
		\draw[very thick] (1,1) -- (2,2);
		\draw[very thick] (1,2) -- (2,2);
		
		\draw[very thick , ->] (2.5,1) -- (3.5,1);
		
		\filldraw (4,0) circle (3pt);
		\filldraw (4,1) circle (3pt);
		\filldraw (4,2) circle (3pt);
		\filldraw (5,0) circle (3pt);
		\filldraw (5,1) circle (3pt);
		\filldraw (5,2) circle (3pt);
		\filldraw (6,0) circle (3pt);
		\filldraw (6,2) circle (3pt);

		\draw[very thick] (4,0) -- (4,1);
		\draw[very thick] (4,1) -- (5,0);
		\draw[very thick] (4,2) -- (5,2);
		\draw[very thick] (5,0) -- (5,1);
		\draw[very thick] (5,1) -- (6,2);
		\draw[very thick] (5,2) -- (6,2);
		\draw[very thick] (6,2) -- (6,0);
		
		\draw[very thick , ->] (6.5,1) -- (7.5,1);
		
		\filldraw (8,0) circle (3pt);
		\filldraw (8,1) circle (3pt);
		\filldraw (8,2) circle (3pt);
		\filldraw (9,0) circle (3pt);
		\filldraw (9,1) circle (3pt);
		\filldraw (10,2) circle (3pt);
		\filldraw (10,0) circle (3pt);

		\draw[very thick] (8,0) -- (8,1);
		\draw[very thick] (8,1) -- (9,0);
		\draw[very thick] (10,0) -- (10,2);
		\draw[very thick] (9,0) -- (9,1);
		\draw[very thick] (9,1) -- (10,2);
		
	\end{tikzpicture}
	\caption{An example of a sequence of one bridge and one break operation.}
	\label{fig:bridge and break chain}
\end{figure}
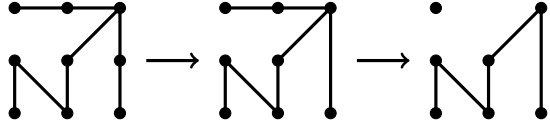

Figure \ref{fig:bridge and break chain} gives an example of multiple applications of the bridge and break operators. We will now define the general states required to solve problem \ref{pt:BIQP problem 2}.

\begin{definition}
	An \emph{extended IQP graph} is represented by $\widetilde{\mathbf{Q}} \in \left\{ -1 , 0 , 1 \right\}^{n_{a} \times n_{p}}$. The vertex set contains $A = \left\{ a_{1} , ... , a_{n_{a}} \right\}$ and $P = \left\{ p_{1} , ... , p_{n_{p}} \right\}$ while $\widetilde{\mathbf{Q}}_{hj}=0$ and $\widetilde{\mathbf{Q}}_{hj}=1$ denote if there is a connection or not between vertices $a_h$ and $p_j$ in the same way as for the IQP graph of Definition \ref{def:IQP graph}.
    
	We interpret $\widetilde{\mathbf{Q}}_{hj}=-1$ as the existence of an intermediary vertex $b_k$ between vertices $p_j$ and $a_{h}$, and denote with $n_b$ the number of -1s in $\widetilde{\mathbf{Q}}$. The vertex set includes the \emph{bridge and break vertices} $B = \left\{ b_{1} , ... , b_{n_b} \right\}$ and the edge set includes edges between $b_{k}$ and $a_{h}$ as well as between $b_{k}$ and $p_{j}$ when $\widetilde{\mathbf{Q}}_{hj}=-1$. We  define a surjective function $g$ for which $g \left( h , j \right) = k $ when $b_{k}$ is the intermediate vertex connected to $a_{h}$ and $p_{j}$. 
\end{definition}

\begin{figure}
	\centering
	\begin{tikzpicture}[scale = 0.7]
		\filldraw (1.5,0) circle (3pt) node[anchor = north east] {$a_{1}$};
		\filldraw (4.5,0) circle (3pt) node[anchor = north east] {$a_{2}$};
		
		\filldraw (0,2) circle (3pt) node[anchor = south east] {$p_{1}$};
		\draw[very thick] (0,2) -- (1.5,0);
		
		\filldraw (0.75,1) circle (3pt) node[anchor = south west] {$b_{1}$};
		
		\filldraw (3,2) circle (3pt) node[anchor = south east] {$p_{2}$};
		\draw[very thick] (3,2) -- (4.5,0);
		
		\filldraw (6,2) circle (3pt) node[anchor = south east] {$p_{3}$};
		\draw[very thick] (6,2) -- (1.5,0);
		
		\draw[very thick] (6,2) -- (4.5,0);
		\filldraw (5.25,1) circle (3pt) node[anchor = north west] {$b_{2}$};
		
		\node at (11,1) {$\widetilde{\mathbf{Q}} =   \left( 
		                                    \begin{array}{ccc}
                                                -1 & 0 & 1 \\
                                                0 & 1 & -1
                                            \end{array}
                                        \right)$};
	\end{tikzpicture}
	\caption{An extended  IQP graph described by $\widetilde{\mathbf{Q}}$ with $(n_a,n_p,n_b)=(2,3,2)$, $P = \left[ p_1 , p_2 , p_3 \right]$, $B = \left[ b_1 , b_2 \right]$ and $A = \left[ a_1 , a_2 \right]$. Two vertices $b_1$ and $b_2$ are introduced and  the function $g: \mathbb{Z}_{n_a\times n_p} \rightarrow \mathbb{Z}_{n_b}$ is defined as $g \left( 1 , 1 \right) = 1$ and $g \left( 2 , 3 \right) = 2$.}
	\label{fig:IQP extended graph}
\end{figure}
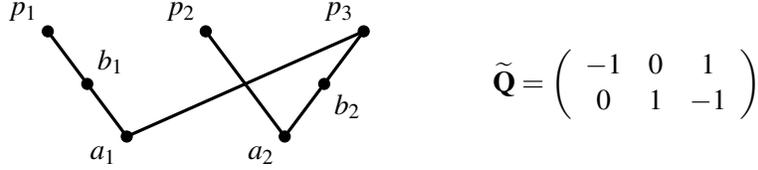

An extended IQP graph $\widetilde{\mathbf{Q}}$ can be built from an IQP graph $\mathbf{Q}$ by replacing any number of the entries of $\mathbf{Q}$ with $-1$. Throughout the remainder of this work we will use tilde notation to represent an extended IQP graph $\widetilde{\mathbf{Q}}$ build from an IQP graph $\mathbf{Q}$ in this way. Figure \ref{fig:IQP extended graph} displays an example of an extended IQP graph. By applying a bridge operator to $b_{1}$ and a break operation to $b_{2}$ in $\widetilde{\mathbf{Q}}$ of Figure \ref{fig:IQP extended graph} we arrive at $\mathbf{Q}$ of Figure \ref{fig:bipartite graph}. It is in this sense that an extended IQP graph is `more general' that an IQP graph.

To solve our three problems we must translate these graph theoretic ideas into operations on quantum states. The following definition allows us to use graphs defined above to describe entanglement patterns.

\begin{definition}
    Consider the extended IQP graph $\widetilde{\mathbf{Q}} \in \left\{-1,0,1\right\}^{n_a\times n_p}$ and use function $g \left( h , j \right) = k$ to define indices $k=1,\dots,n_b$ for the elements $\widetilde{\mathbf{Q}}_{hj}=-1$. The circuit $E_{\widetilde{\mathbf{Q}}}$ on $(n_a + n_p +n_b)$ qubits applies controlled-$Z$ operations between qubits $p_{j}$ and $a_{h}$ if $\widetilde{\mathbf{Q}}_{hj} = 1$ and, between qubits $b_{g(h,j)}$ and $a_{h}$, and, $b_{g(h,j)}$ and $p_{j}$, when $\widetilde{\mathbf{Q}}_{hj} = -1$.
\end{definition}

Now we reformulate a lemma from \cite{fitzsimons2012unconditionally} in order to translate bridge and break operations from graph theoretical ideas into operations on quantum states.

\begin{lemma}
	\label{lem:state bridge and break correctness}
	Consider a quantum state $E_{\mathbf{Q}}\ket{\phi}$ where $\ket{\phi}$ is arbitrary. If $\widetilde{\mathbf{Q}}$ is an extended IQP graph built from $\mathbf{Q}$ then there exists a state $E_{\widetilde{\mathbf{Q}}}\ket{\psi}$, which can be transformed into the state $E_{\mathbf{Q}}\ket{\phi}$ through a sequence of Pauli-$Y$ basis measurements on qubits and local rotations around the Z axis on the unmeasured qubits through angles $\left\{ 0 , \frac{\pi}{2} , \pi , \frac{3 \pi}{2} \right\}$.
\end{lemma}

The detailed proof of Lemma \ref{lem:state bridge and break correctness} \cite{mills2017information} shows us that we can create the following state where $p_j$ and $a_{h}$ are the primary and ancillary qubits connected to $b_k$ (i.e. $g \left( h , j \right) = k$), $r^{b}$ and $d^{b}$ are strings described later in Algorithm \ref{alg:real IQP resource honest server} and $s_k^{b}$ is the outcome of a measurement on qubit $b_{k}$.

\begin{equation}
	\label{equ:section final state}
	\prod_{k = 1}^{n_{b}} \left( S_{p_{j}}^{(-1)^{s^{b}_{k}+r^{b}_{k}}} \otimes S_{a_{h} }^{(-1)^{s^{b}_{k}+r^{b}_{k}}} \right)^{d^{b}_{k}} \left( Z_{p_{j}}^{r^{b}_{k}} \otimes Z_{a_{j}}^{r^{b}_{k}} \right)^{1-d^{b}_{k}} E_{\mathbf{Q}}\ket{\phi}
\end{equation}

To achieve this we make measurements of the qubits corresponding to bridge and break vertices (which we call \emph{bridge and break qubits} and which are initialised as $\ket{b_k} = Y^{r^{b}_{k}} \sqrt{Y}^{d^{b}_{k}} \ket{0}$) in $E_{\widetilde{\mathbf{Q}}}\ket{\psi}$ in the Pauli $Y$ basis. Using this method, we generate the state in Lemma \ref{lem:IQP graph design} (i.e. $E_{\mathbf{Q}} \bigotimes_{1}^{n_{a} + n_{p}} \ket{+}$) up to $S$ corrections, which can be accounted for by correcting the primary and ancillary measurement bases.


\subsection{Blindness}
\label{sec:security proof}

\begin{figure}
	\centering
	\begin{tikzpicture}[scale = 0.7]
     
		\draw[very thick] (0,0.5) rectangle (8,2.5) node[align = center, pos = 0.5] {$\widetilde{x} = \begin{cases} x & \text{ if honest} \\ \mathcal{E} \left( \mathbf{Q} , \rho_{B} , \theta \right) & \text{if dishonest}\end{cases}$} node[anchor = south east] {$\mathcal{S}$};
    
		\draw[very thick, ->] (-2,2) -- (0,2) node[anchor = south, pos = 0.5] {$\mathbf{Q}$};
		\draw[very thick, <-] (-2,1) -- (0,1) node[anchor = south, pos = 0.5] {$\widetilde{x}$};
    
		\draw[very thick, dotted, <-] (8,2) -- (10,2) node[anchor = south, pos = 0.5] {$\mathcal{E}$};
		\draw[very thick, dotted, <-] (8,1) -- (10,1) node[anchor = south, pos = 0.5] {$\rho_{B}$};
    
		\draw[very thick, <->] (-2,0) -- (10,0) node[anchor = north, pos = 0.5] {$\widetilde{\mathbf{Q}} , \mathcal{Q},\theta$};
	\end{tikzpicture}
	\caption{The ideal blind delegated IQP computation resource.}
	\label{fig:ideal resource}
\end{figure}
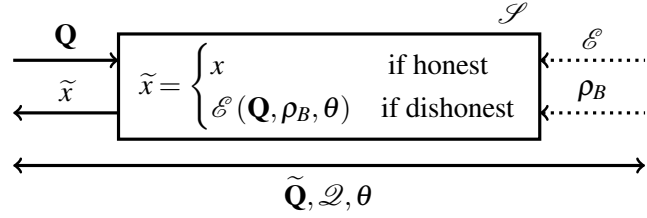

To address problem \ref{pt:BIQP problem 3}, we wish to construct the \emph{Ideal Resource} of Figure \ref{fig:ideal resource} which takes as input from the client an IQP computation $\left( \mathbf{Q} , \theta \right)$, and in return gives a classical output $\widetilde{x}$. If the Server is honest, $\widetilde{x}$ comes from the distribution corresponding to $\left( \mathbf{Q} , \theta \right)$. If the Server is dishonest, they can input some quantum operation $\mathcal{E}$ and some quantum state $\rho_{B}$ and force the output to the client into the classical state $\mathcal{E}\left( \mathbf{Q} , \theta , \rho_{B} \right)$. We assume that the following are public knowledge: an extended IQP graph $\widetilde{\mathbf{Q}}$, the distribution $\mathcal{Q}$ over the possible $\mathbf{Q}$s from which $\widetilde{\mathbf{Q}}$ could be built, and $\theta$. It is important to note that neither $\mathbf{Q}$, nor anything besides that which we have specified as public, is revealed to the Server.

\begin{figure}
	\centering
	\begin{tikzpicture}[scale = 0.7]
		\draw[very thick] (0,0.5) rectangle (1,4.5) node[anchor = south east] {$\pi_{A}$};
    
		\draw[very thick, ->] (-2,4) -- (0,4) node[anchor = south, pos = 0.5] {$\mathbf{Q}$};
		\draw[very thick, <-] (-2,1) -- (0,1) node[anchor = south, pos = 0.5] {$\widetilde{x}$};
    
		\draw[very thick, ->] (1,4) -- (5,4) node[anchor = south, pos = 0.5] {$\rho$};
		\draw[very thick, <-] (1,3) -- (5,3) node[anchor = south, pos = 0.5] {$s^b$};
		\draw[very thick, ->] (1,2) -- (5,2) node[anchor = south, pos = 0.5] {$A,\Pi$};
		\draw[very thick, <-] (1,1) -- (5,1) node[anchor = south, pos = 0.5] {$s^a,s^p$};
    
		\draw[very thick] (5,0.5) rectangle (6,4.5) node[anchor = south east] {$\pi_{B}$};
    
		\draw[very thick, dotted, <-] (6,4) -- (8,4) node[anchor = south, pos = 0.5] {$\mathcal{E}$};
		\draw[very thick, dotted, <-] (6,3) -- (8,3) node[anchor = south, pos = 0.5] {$\rho_{B}$};
		\draw[very thick, dotted, ->] (6,1) -- (8,1) node[anchor = south, pos = 0.5] {$\rho_{B}'$};
    
		\draw[very thick, <->] (-2,0) -- (8,0) node[anchor = north, pos = 0.5] {$\widetilde{\mathbf{Q}} , \mathcal{Q}, \theta$};
		
		\draw[very thick, dotted] (1.5,0.5) rectangle (4.5,5) node[anchor = south east] {$\mathcal{R}$};
    
	\end{tikzpicture}
	\caption{The real communication protocol. $\pi_{A}$ is the set of operation performed by the client, $\pi_{B}$ are those of the Server and $\mathcal{R}$ is the communication channel (quantum and classical) used by the client and the Server in the protocol. Details of the communication protocol $\mathcal{R}$ can be seen in Algorithm \ref{alg:real IQP resource honest server}.}
	\label{fig:real IQP computation resource}
\end{figure}
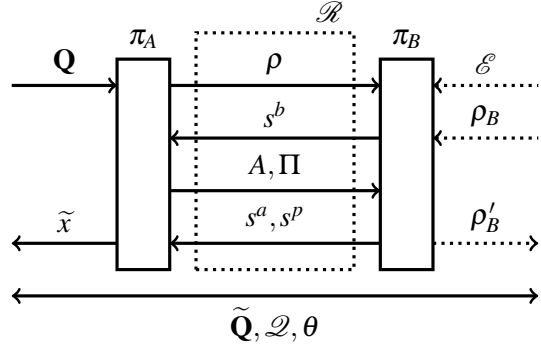

Blindness is added to the work of Section \ref{sec:Break, Bridge Operators} by performing random rotations when initialising the primary and ancillary qubits. These are corrected by rotating the measurement bases of those qubits, therefore ensuring the original IQP computation is performed. Intuitively, this randomness provides blindness as it hides the previous corrections shown in equation \eqref{equ:section final state}, which would otherwise give away if a bridge or break operation was applied to a neighbour. Our communication protocol can be seen in Algorithm \ref{alg:real IQP resource honest server} and is summarised in Figure \ref{fig:real IQP computation resource}.

\begin{algorithm}
	\caption{Blind distributed IQP computation}
	\label{alg:real IQP resource honest server}
	\textbf{Public:} $\widetilde{\mathbf{Q}} , \mathcal{Q}$, $\theta$
  
	\textbf{client input:} $\mathbf{Q}$ 
  
	\textbf{client output:} $\widetilde{x}$ 
  
	\textbf{Protocol:}
  
	\begin{algorithmic}[1]
		\STATE The client randomly generates $r^{p} , d^{p} \in \left\{ 0 , 1 \right\}^{n_{p}}$ and $r^{a} , d^{a} \in \left\{ 0 , 1 \right\}^{n_{a}}$ where $n_{p}$ and $n_{a}$ are the numbers of primary and ancillary qubits respectively.
		\label{alg line:real IQP resource honest server - primary and ancila random key generation}
		\STATE The client generates the states $\ket{p_{j}} = Z^{r^{p}_{j}} S^{d^{p}_{j}} \ket{+} $ and $\ket{a_{h}}  =  Z^{r^{a}_{h}} S^{d^{a}_{h}} \ket{+}$ for $j \in \left\{ 1 , \dots, n_{p} \right\}$ and $h \in \left\{ 1 , \dots, n_{a} \right\}$.
		\label{alg line:real IQP resource honest server - primary and ancillary state generation}
		\STATE The client creates $d^b \in \left\{0,1\right\}^{n_b}$ in the following way: For $h=1,\dots,n_a$ and $j=1,\dots,n_p$, if $\widetilde{\mathbf{Q}}_{hj}=-1$ and $\mathbf{Q}_{hj}=0$, then $d^b_k=0$ else if $\widetilde{\mathbf{Q}}_{hj}=-1$ and $\mathbf{Q}_{hj}=1$ then $d^b_k=1$. He keeps track of the relation between $k$ and $(h,j)$ via the surjective function $g: \mathbb{Z}_{n_a \times n_p} \rightarrow \mathbb{Z}_{n_b}$.  
		\STATE The client generates $r^{b} \in \left\{ 0 , 1 \right\}^{n_{b}}$ at random and produces the states $\ket{b_{k}} = Y^{r^{b}_{k}} \left( \sqrt{Y} \right)^{d^{b}_{k}} \ket{0} $ for $k \in \left\{ 1 , \dots, n_{b} \right\}$.
		\label{alg line:real IQP resource honest server - bridge and break state generation}
		\STATE The state $\rho$, comprising of all of the client's produced states, is sent to the Server.
		\STATE The Server implements $E_{\widetilde{\mathbf{Q}}}$.
		\STATE The Server measures qubits $b_{1} , \dots, b_{n_{b}}$ in the $Y$-basis $\left\{ \ket{+^{Y}} , \ket{-^{Y}} \right\}$ and sends the outcome $s^b \in \left\{ 0 , 1 \right\}^{n_{b}}$ to the client.
		\STATE The client calculates $\Pi^{z} , \Pi^{s} \in \left\{ 0 , 1 \right\}^{n_{p}}$ and $A^{z} , A^{s} \in \left\{ 0 , 1 \right\}^{n_{a}}$ using equations \eqref{equ:primary Z correction term}, \eqref{equ:primary S correction term}, \eqref{equ:ancila Z correction term} and \eqref{equ:ancila S correction term}. 
		\begin{align}
			\label{equ:primary Z correction term}
			\Pi^{z}_{j} &= \sum_{h,k:g(h,j)=k} r_k^b \left( 1 - d^{b}_k \right) - r^{p}_{j} \\
			\label{equ:primary S correction term}
			\Pi^{s}_{j} &= \sum_{h,k:g(h,j)=k} (-1)^{s^{b}_k+r^{b}_k} d^{b}_k  - d^{p}_{j}\\
			\label{equ:ancila Z correction term}
			A^{z}_{h} &= \sum_{j,k:g(h,j)=k} r_k^{b} \left( 1 - d^{b}_k \right) - r^{a}_{h}\\
			\label{equ:ancila S correction term}
			A^{s}_{h} &= \sum_{j,k:g(h,j)=k}(-1)^{s^{b}_k+r^{b}_k}  d^{b}_k  - d^{a}_{h}
		\end{align}
		\STATE The client sends, to the serve, $A \in \left\{0,1,2,3\right\}^{n_a}$ and $\Pi \in \left\{0,1,2,3\right\}^{n_p}$ for the ancillary and primary qubits respectively, where $A_{h} = A^{s}_{h} + 2 A^{z}_{h} \pmod 4$ and $\Pi_{j} = \Pi^{s}_{j} + 2 \Pi^{z}_{j} \pmod 4$.
		\STATE The Server measures the respective qubits in the basis below for the ancillary and primary qubits respectively. 
		\begin{equation}
			\label{equ:primary and ancillary measurement basis}
			S^{- A_{h} } \left\{ \ket{0_\theta} , \ket{1_\theta} \right\}  \text{ and } S^{- \Pi_{j} } \left\{ \ket{+} , \ket{-} \right\}
		\end{equation}
		\STATE The measurement outcomes $s^{p} \in \left\{ 0 , 1 \right\}^{n_{p}}$ and $s^{a} \in \left\{ 0 , 1 \right\}^{n_{a}}$ are sent to the client.
		\STATE The client generates and outputs $\widetilde{x} \in \left\{ 0 , 1 \right\}^{n_{p}}$ as follows. 
		\begin{equation}
			\label{equ:IQP final outcome calculation}
			\widetilde{x}_{j} = s^{p}_{j} + \sum_{h:\mathbf{Q}_{hj} = 1}  s^{a}_{h} \pmod 2
		\end{equation}
	\end{algorithmic}
 
\end{algorithm}

To prove composable security of the proposed protocol we drop the notion of a malicious Server for that of a global distinguisher which has a view of all inputs and outputs of the relevant resources. To be able to compare the ideal resource with the real communication protocol, we develop a simulator $\sigma$ interfacing between the ideal resource $\mathcal{S}$ of Figure \ref{fig:ideal resource} and the distinguisher such that the latter cannot tell the difference between an interaction with the ideal resource and the real protocol. We employ the Abstract Cryptography framework introduced in \cite{maurer2011abstract,portmann2014cryptographic} and teleportation techniques inspired by \cite{dunjko2014composable} to prove security in the case of a malicious Server. If $\pi_{A}$ is protocol run by the Client and $\mathcal{R}$ is the communication channel between them and the Server, then we can prove \cite{mills2017information} that:

\begin{equation}
	\pi_A\mathcal{R}\equiv \mathcal{S}\sigma
\end{equation}
where equality here means the two protocols are indistinguishable.

\begin{theorem}
    \label{thm:security proof}
    The communication protocol described by Algorithm \ref{alg:real IQP resource honest server} is information theoretically secure against a dishonest Server. 
\end{theorem}

We can now be sure that our communication protocol is indistinguishable from an ideal resource which performs an IQP computation without communicating any information to the Server which is not already public. This is proven in a composable framework \cite{portmann2014cryptographic,dunjko2014composable} and so can be used as part of future protocols, as we will do in section \ref{sec:hypothesis test}.


\section{The Hypothesis Test}
\label{sec:hypothesis test}

\subsection{Previous work}
\label{sec:SB hypothesis test}

The general idea of our \emph{Hypothesis Test}, building on the work of \cite{shepherd2009temporally}, is that there is some hidden structure in the program elements, $\mathbf{q}_{h}$, of the $X$-program, which results in some structure in the distribution of the outputs, known only to the client. The client can use this structure to check the Server's reply. A Server possessing an IQP machine can reproduce this structure by implementing the $X$-program. A Server not in possession of an IQP machine, cannot generate outputs obeying the same rules. We summarise this discussion by three conditions that a hypothesis test using this method must meet.
\begin{enumerate}[label = \arabic{list}.\arabic*]
	\item \label{pt:hypothesis test IQP hard} A client asks a Server to perform a hard-to-classically-simulate IQP computation. 
	\item \label{pt:hypothesis test structure} The client can check the solution to this computation because they know some secret structure that makes this checking process efficient.
	\item \label{pt:hypothesis test hidden structure} The Server must be unable to uncover this structure in polynomial time.
\end{enumerate}
\stepcounter{list}
The particular `known structure' of the output used in \cite{shepherd2009temporally} to satisfy \ref{pt:hypothesis test structure} is its \emph{bias}. 

\begin{definition}
	\label{def:bias}
	If $\mathtt{X}$ is a random variable taking values in $\left\{ 0 , 1 \right\}^{n_p}$ and $\mathbf{s} \in \left\{ 0 , 1 \right\}^{n_p}$ then the bias of $\mathtt{X}$ in the direction $\mathbf{s}$ is $\mathbb{P} \left( \mathtt{X} \cdot \mathbf{s}^T = 0 \right)$ where the product is performed modulo 2. Hence, the bias of a distribution in the direction $\mathbf{s}$ is the probability of a sample from the distribution being orthogonal to $\mathbf{s}$.
\end{definition}

To calculate the bias of $\mathtt{X}$ in direction $\mathbf{s} \in \left\{ 0 , 1 \right\}^{n}$, we form the linear code $\mathcal{C}_{\mathbf{s}}$ by selecting all rows, $\mathbf{q}_{h}$, of the $X$-program, $\left(\mathbf{Q}, \theta \right) \in \left\{ 0 , 1 \right\}^{n_{a} \times n_{p}} \times \left[ 0 , 2 \pi \right]$, such that $\mathbf{q}_{h} \cdot \bf{s}^{T} = 1$. We form, from them, the matrix, $\mathbf{Q}_{\mathbf{s}}$, which is set as the generator matrix of $\mathcal{C}_{\bf{s}}$. Defining $n_{\bf{s}}$ as the number of rows of $\mathbf{Q}_{\bf{s}}$ and $\# \mathbf{c}$ as the hamming weight of $\mathbf{c}$ allows us to understand the following expression derived in \cite{shepherd2009temporally}.
\begin{equation}
	\label{equ:bias prediction}
	\mathbb{P} \left( \mathtt{X} \cdot \mathbf{s}^{T} = 0 \right) = \mathbb{E}_{ \mathbf{c} \sim \mathcal{C}_{\mathbf{s}}} \left[ \cos^{2} \left( \theta \left( n_s - 2 \cdot \# \mathbf{c} \right) \right) \right]
\end{equation}

Hence, the bias of an $X$-program in the direction $\bf{s}$ depends only on $\theta$ and the linear code defined by the generator matrix $\mathbf{Q}_{\mathbf{s}}$. One can now imagine a hypothesis test derived from this. Although the $X$-program to be implemented needs to be made public, the direction $\mathbf{s}$ which will be used for checking, will be kept secret. This gives a client, with the computational power to calculate the quantity of expression \eqref{equ:bias prediction}, the necessary information to compute the bias, but does not afford the Server the same privilege. 

What we want to show is that the only way for the Server to produce an output with the correct bias is to use an IQP machine. If the Server could uncover $\mathbf{s}$ then they could calculate the value of expression \eqref{equ:bias prediction} and return vectors to the client which are orthogonal to $\mathbf{s}$ with the correct probability. We specialise the conditions mentioned at the beginning of this section to this particular method. 

\begin{enumerate}[label = \arabic{list}.\arabic*]
	\item \label{pt:bias hypothesis test IQP hard} The $X$-Program sent to a Server represents an IQP computation that is hard to classically simulate.
	\item \label{pt:bias hypothesis test structure} It must be possible for a client, having knowledge of a secret $\mathbf{s}$ and the $X$-program, to calculate the quantity of expression \eqref{equ:bias prediction}.
	\item \label{pt:bias hypothesis test hidden structure} The knowledge of the Server must be insufficient to learn the value of $\bf{s}$.
\end{enumerate}
\stepcounter{list}

In \cite{shepherd2009temporally} the authors develop a protocol for building an $X$-program and a vector $\mathbf{s}$ performing this type of hypothesis test. The code $\mathcal{C}_{\mathbf{s}}$ used is a quadratic residue code with $\theta = \frac{\pi}{8}$ and condition \ref{pt:bias hypothesis test IQP hard} is conjectured to be satisfied by $X$-program matrices generating this code space. This conjecture is supported by giving a classical simulation that is believed to be optimal and achieves maximum bias value $0.75$, lower than that expected from an IQP machine. Condition \ref{pt:bias hypothesis test structure} is satisfied in \cite{shepherd2009temporally}, by explicitly calculating the bias value to be $\cos^{2}\left( \frac{\pi}{8} \right)$ for their choice of $X$-program and $\mathbf{s}$. 

The way in which condition \ref{pt:bias hypothesis test hidden structure} is addressed in \cite{shepherd2009temporally} relies on the fact that the right-hand side of equation \eqref{equ:bias prediction} is equal for all generator matrices in a \emph{matroid} \cite{oxley2006matroid}.

\begin{definition}
	A $h$-point binary \emph{matroid} is an equivalence class of matrices with $h$ rows, defined over $\mathbb{F}_2$. Two matrices, $\mathbf{M}_1$ and $\mathbf{M}_2$, are said to be equivalent if, for some permutation matrix $\mathbf{R}$, the column echelon reduced form of $\mathbf{M}_1$ is the same as the column echelon reduced form of $\mathbf{R} \cdot \mathbf{M}_2$ (In the case where the column dimensions do not match, we define equivalence  by deleting columns containing only $0$s after column echelon reduction).
\end{definition}

Consider right-multiplying the matrix $\mathbf{A}$ on $\mathbf{Q}$ where $\mathbf{A}$ is chosen so that $\mathbf{Q} \mathbf{A}$ belongs to the same matroid as $\mathbf{Q}$. Notice that $\mathbf{q}_{h} \mathbf{s'}^{T} = \left( \mathbf{q}_{h} \mathbf{A} \right) \left( \mathbf{A}^{-1} \mathbf{s'}^{T} \right)$ and so rows originally non-orthogonal to $\mathbf{s'}^{T}$ are now non-orthogonal to $\mathbf{A}^{-1} \mathbf{s'}^{T}$. Hence, we can locate the matroid $\mathbf{Q}_{\mathbf{s'}}$ in the matrix $\mathbf{Q} \mathbf{A}$ by using $\mathbf{A}^{-1} \mathbf{s'}^{T}$. We now understand what to do to the $\mathbf{Q}$ describing the $X$-program we are considering, so that we remain in the same matroid, and hence prevent the bias from changing, if we perform operations on $\mathbf{s'}$.

To hide the full description of $\mathbf{Q}$, use instead $\mathbf{Q}'$, built by adding additional random rows orthogonal to $\mathbf{s'}$, which do not affect the value of the bias. The combination of matrix randomisation, and the the fact that they do not know $\mathbf{Q}'$, makes it hard, as conjectured in \cite{shepherd2009temporally}, up to some computational complexity assumptions, for the Server to recover $\mathbf{A}^{-1} \mathbf{s'}^{T}$ if they are given $\mathbf{Q}' \mathbf{A}$, even when they know $\mathbf{s'}$ and $\mathbf{Q}$. It is now a matter for the Server to implement the $X$-program $\mathbf{Q}' \mathbf{A}$ and for the client to check the bias of the output in the hidden direction $\mathbf{s}^{T} = \mathbf{A}^{-1} \mathbf{s'}^{T}$. This is the idea behind the approach used by \cite{shepherd2009temporally} to address condition \ref{pt:bias hypothesis test hidden structure}.


\subsection{Our Protocol}
\label{sec:our protocol}
The main contribution of this work is to revisit condition \ref{pt:bias hypothesis test hidden structure}.

\begin{theorem}
    Algorithm \ref{alg:Our hypothesis test protocol} presents an information-theoretically secure solution to condition \ref{pt:bias hypothesis test hidden structure}.
\end{theorem}

In Algorithm \ref{alg:Our hypothesis test protocol} we provide a hypothesis test that uses Algorithm \ref{alg:real IQP resource honest server} to verify quantum superiority. By using the blind IQP computation resource of Section \ref{sec:security proof} we have solved condition \ref{pt:bias hypothesis test hidden structure} but do so now with information theoretic security as opposed to the reliance on computational complexity assumptions used by \cite{shepherd2009temporally}. This is true because the Server learns only the distribution $\mathcal{Q}$ over the possible set of graphs $\mathbf{Q}$. By setting $\mathbf{Q} = \mathbf{Q_{s}} \mathbf{A}$,  Algorithm \ref{alg:Our hypothesis test protocol} develops a bijection mapping $\widehat{\mathbf{s}} \in \{0,1\}^{n_{p} - 1}$ to a unique matrix $\mathbf{Q} \in \{0,1\}^{n_{a} \times n_{p}}$. So $\mathcal{Q}$ is equivalent to the distribution from which $\widehat{\mathbf{s}}$ is drawn. In this case it is the uniform distribution over a set of size $2^{n_{p} - 1}$. In particular the server does not learn $\mathbf{s}$ as is required by condition \ref{pt:bias hypothesis test hidden structure}. Notice that Algorithm \ref{alg:Our hypothesis test protocol} should be repeated to accurately check the bias is $\cos^2 \left( \frac{\pi}{8}  \right)$ as we would expect from the quantum case.

\begin{algorithm}[ht]
    \caption{Our hypothesis test protocol}
    \label{alg:Our hypothesis test protocol}  
    \textbf{Input:} $n_{a}$ prime such that $n_{a} + 1$ is a multiple of $8$.
    
    \textbf{client output:} $o \in \left\{ 0 , 1 \right\}$ 
  
    \textbf{Protocol:}
  
    \begin{algorithmic}[1]
        \STATE Set $n_{p} = \frac{n_{a} + 1}{2}$
        \STATE \label{alg line: Our hypothesis test protocol - start with QR} Take the quadratic residue code generator matrix $\mathbf{Q_{r}} \in \left\{ 0 , 1 \right\}^{n_{a} \times \left( n_{p} - 1 \right)}$ 
        \STATE \label{alg line: Our hypothesis test protocol - add column of ones}Let $\mathbf{Q_{s}} \in \left\{ 0 , 1 \right\}^{n_{a} \times n_{p}}$ be $\mathbf{Q_{r}}$ with a column of ones appended to the last column. 
        \STATE Pick $\widehat{\mathbf{s}} \in \left\{ 0 , 1 \right\}^{n_p - 1}$ chosen uniformly at random. 
        \STATE \label{alg line: Our hypothesis test protocol - randomisation matrix} Define the matrix $\mathbf{A} \in \left\{ 0 , 1 \right\}^{n_p \times n_p}$ according to equation \eqref{equ:transformation matrix}.

        \begin{equation}
            \label{equ:transformation matrix}
            \mathbf{A}_{h,j} =  \begin{cases} 
                                    1 & \text{if } h = j \\ 
                                    0 & \text{if } h \neq j \text{ and } j < n_{p} \\ 
                                    \widehat{\mathbf{s}}_{h} & \text{if } j = n_{p} \text{ and } h < n_p
                                \end{cases}
        \end{equation}
 
        \STATE Set $\mathbf{Q} = \mathbf{Q_{s}} \mathbf{A}$ and $\theta = \frac{\pi}{8}$.
        \STATE Set $\widetilde{\mathbf{Q}}$ to be the matrix $\mathbf{Q_{r}}$ with a column of $-1$ appended.
        \STATE Set $\mathcal{Q}$ to be the uniform distribution over all possible $\mathbf{Q}$ for different $\widehat{\mathbf{s}}$.
        \STATE Perform IQP computation $\mathbf{Q}$ using Algorithm \ref{alg:real IQP resource honest server} with inputs $\mathbf{Q}$, $\widetilde{\mathbf{Q}}$, $\mathcal{Q}$ and $\theta$ and outputs $\widetilde{x}$ and  $\rho_{B}'$.
        \STATE Let $\mathbf{s'} \in \left\{ 0 , 1 \right\}^{n_p}$ be the vector with entries all equal to zero with the exception of the last which is set to one.
        \STATE \label{alg line:Our hypothesis test protocol - test orthogonality} Test the orthogonality of the output $\widetilde{x}$ against $\mathbf{s}^{T} = A^{-1} \mathbf{s'}^{T}$ setting $o = 0$ if it is not orthogonal and $o = 1$ if it is orthogonal.
    \end{algorithmic}
 
\end{algorithm}


\section{Discussion, Conclusion and Future Work}

We have presented a new certification technique for IQP machines which can be run by a client able to prepare single-qubit Pauli eigenstates. By giving the client minimal quantum abilities we can remove computational restrictions placed on the Server in previous work \cite{shepherd2009temporally} and, instead, prove information-theoretical security against an untrusted Server. 

There are several advantages of using this tailored verification protocol for IQP computations, rather than a straightforward verification in a universal quantum computing model \cite{fitzsimons2012unconditionally,kashefi2017optimised}. The latter requires higher precision in the manipulation of single qubits from the client (use of states other than single-qubit Pauli eigenstates) and significantly more quantum communication and processing for the verification technique. Although the qubit consumption is still, as in this work, linear in the size of the computation, in early machines the constant factor will likely be important. Further, asking a Server to perform an IQP computation using a model that is universal for quantum computation would require the Server to create large cluster states and perform measurement that might lie far beyond its IQP capabilities. 

IQP circuits are important as they may prove easier to implement experimentally compared to universal quantum computers. Due to the commutativity of the gates it is theoretically possible to perform an IQP computation in one round of measurements. Our protocol requires a two-round MBQC computation which we believe is not a significant additional requirement, given the implementation requirements of universal quantum computation. We therefore believe that our scheme will be implementable in the near future, for a small number of qubits, and so a future avenue of research would be to study this hypothesis test protocol under realistic experimental errors, following similar examples of work in this direction \cite{bremner2016achieving,kapourniotis2017nonadaptive}. 

The demand for the Server to have memory to support a two-round MBQC computation means machines capable of passing the original test of \cite{shepherd2009temporally} might not be able to pass that of this work. These studies do not restrict the architectures that the Server can use, which comes at the high cost of placing computational restrictions on the Server \cite{shepherd2009temporally}. Our requirement that the Server can perform two-round MBQC allows us to achieve information-theoretical security. Finally, given that Gaussian quantum subtheory can be seen as a continuous variable analogue of the stabiliser formalism \cite{spekkens2016quasi,bartlett2002efficient,bartlett2012reconstruction}, a natural extension would be a continuous variable analogue of our protocol where the client prepares only Gaussian states.


\section{Acknowledgements}

The authors would like to thank Andru Gheorghiu and Petros Wallden for enlightening discussions and feedback. This work was supported by grant EP/L01503X/1 for the University of Edinburgh School of Informatics Centre for Doctoral Training in Pervasive Parallelism, from the UK Engineering and Physical Sciences Research Council (EPSRC) and by grants  EP/K04057X/2, EP/N003829/1 and EP/ M013243/1, as well as by the European Union’s Horizon 2020 Research and Innovation program under Marie Sklodowska-Curie Grant Agreement No. 705194.


\bibliographystyle{eptcs}
\bibliography{bibliography}

\end{document}